%% file: HHirabayashi.tex
\documentclass{evn2004}

\usepackage{txfonts}
\usepackage{graphicx}

\authorrunning{Hirabayashi et al.}
\titlerunning{The VSOP-2 Mission}

\begin{document}
   \title{On the Near-term Space VLBI Mission VSOP-2}

   \author{H. Hirabayashi\inst{1}, Y. Murata\inst{1}, P.G. Edwards\inst{1}
           Y.~Asaki\inst{1}, N.~Mochizuki\inst{1}, M.~Inoue\inst{2},
           T.~Umemoto\inst{2}, S.~Kameno\inst{2} \and Y.~Kono\inst{2}
          }

   \institute{Institute of Space and Astronautical Science (ISAS)/JAXA, Sagamihara, Kanagawa 229-8510, Japan
         \and
              National Astronomical Observatory of Japan, Mitaka, Tokyo 181-8588, Japan
             }

   \abstract{
A second generation near-term space VLBI mission,
VSOP-2, is being planned for a launch in 2010 or soon after.  The
scientific objectives are very high angular resolution imaging of
astrophysically exotic regions, including the cores, jets, and accretion 
disks of active galactic nuclei (AGN), water maser emissions,
micro-quasars, coronae of young stellar objects, etc.  A highest
angular resolution of about 
40\,$\mu$as 
is achieved in the 
43\,GHz band. Engineering developments are in progress for the deployable
antenna, antenna pointing, high data rate transmission, cryogenic receivers, 
accurate orbit determination, etc., to realize this
mission. International collaboration will be as important as it has been
for VSOP.

   }

   \maketitle
%

\section{Introduction}

Following the successes of the VLBI Space Observatory Programme
(Hirabayashi et al. \cite{hir98,hir00}), 
a near-term next generation Japanese
space VLBI mission, currently referred to as VSOP-2 
(Hirabayashi et al. \cite{hir01}; Hirabayashi \cite{hir04}), 
is being planned 
in collaboration with international partners.  
The VSOP-2 website is
http://www.vsop.isas.jaxa.jp/vsop2/ .

\section{VSOP-2 Science Goals}

The VSOP-2
science goals include: study of emission mechanisms in conjunction
with the next generation of X-ray and gamma-ray satellites; full
polarization studies of magnetic field orientation and evolution in
jets, and measurements of Faraday rotation towards AGN cores; high
linear resolution observations of nearby AGN to probe the formation
and collimation of jets and the environment around supermassive black
holes (e.g., Figure~1); and the highest resolution studies of spectral line masers and
mega-masers, and circum-nuclear disks.  

The VSOP mission has made important contributions to the study
of jets (e.g., Piner et al.\ \cite{pin}; Lobanov \& Zensus \cite{lob}; 
Murphy et al.\ \cite{mur})
and plasma torii (Jones et al.\ \cite{jon}; Kameno et al.\ \cite{kam}), 
but the VSOP-2 mission    
can pin point the more fundamental inner region by its higher observing        
frequency and by higher angular resolution.  In the nearby AGN M87, the 
40$\mu$as beam can discern the relative locations of the 
black hole, accretion disk, and postulated shock regions.

Magnetic field structures of young stellar objects by non-thermal 
gyro-synchrotron emissions is a new area to be studied in detail
(see Figure~2).
The X-ray satellites TENMA and GINGA detected hard X-ray emissions from
molecular clouds (Koyama 1987, Koyama et al.\ 1992), and ASCA detected
quasi-periodic flare from the proto-star YLW15A (Tsuboi et al.\ 2000),
suggesting reconnection of magnetic fields connecting the star and its
accretion disk. Such a new scientific area could be 
investigated by VSOP-2, for example.

By both the superb angular resolution and magnetic field information, 
theory and magneto-hydrodynamic simulations can be guided and tested 
in these astronomical objects.

   \begin{figure}
   \centering
   \vspace{300pt}
   \includegraphics{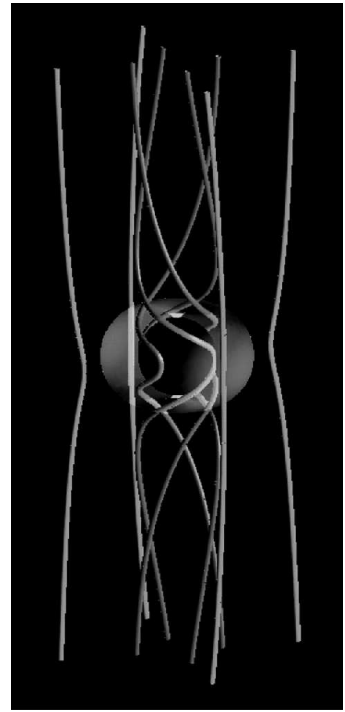}
      \caption{Magneto-Hydro-Dynamic (MHD)
simulation of the magnetic field structure around a black hole (from Koide et al.\ \cite{koi}).
         \label{fig1}
         }
   \end{figure}

  \begin{figure*}
   \centering
   \vspace{220pt}
   \includegraphics{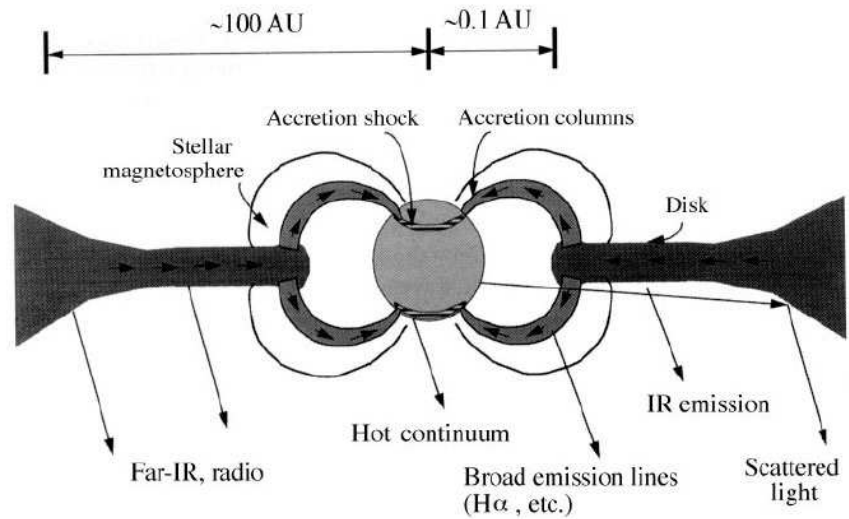}
      \caption{Schematic diagram of the scales and processes
in young stellar objects (from Hartman \cite{har98}).
              }
         \label{f2}
   \end{figure*}


\section{VSOP-2 Instrumental Capabilities}

To realize the near-term space-VLBI mission VSOP-2 and make it scientifically 
valuable, we have three basic 
improvements over VSOP as follows: 
\begin{enumerate}{}
\item an order of magnitude increase in the maximum 
observing frequency, from 5\,GHz to 43\,GHz, allowing detailed imaging deeper 
into the emitting plasma 
\item an order of magnitude increase of maximum angular 
resolution, to $\sim$40\,$\mu$as, using the above frequency and 
a 25,000\,km apogee height. This apogee height with 1,000\,km 
perigee height is selected from the     
consideration of good imaging orbit with the ground array. 
\item an order of magnitude increase in interferometer sensitivity for 
continuum observations.
This is accomplished mainly by the 8 times higher bit rate and the 
lower system noise temperatures.          
\end{enumerate}

These are shown quantitatively in Table~1. 
The observing bands will be 8, 22, and 43\,GHz and 
the receivers for the 22 and 43~GHz bands will be 
cooled.

VSOP-2 can receive both sense of polarizations (LHCP and RHCP)
simultaneously within the limitation of 1~Gbps data transfer. This
will enhance polarization mapping capability. HALCA, on the other
hand, received and transferred only LHCP, but scientific observations
were successfully performed using simultaneous LHCP
and RHCP ground observations (e.g., Kemball et al.\ 2000;
Gabuzda \& G{\' o}mez 2001; Gabuzda 2003).

Observations at 22\,GHz were planned with HALCA, but due (most likely)
to damage to the wavelguide connection during launch, open
scientific observations were not possible. Water vapour maser
and mega-maser studies with space VLBI resolution thus remain a
new area for VSOP-2.

\section{The VSOP-2 Satellite}
The VSOP-2 spacecraft will  employ a
9\,m off-axis paraboloid antenna.  
It is assumed the VSOP-2 satellite will be
launched on a M-V rocket and placed
in an elliptical orbit with an apogee height of $\sim$25,000\,km and a
perigee height of $\sim$1,000\,km, resulting in a period of
$\sim$7.5~hours. Unlike HALCA, the VSOP-2 satellite will receive both LCP
and RCP, and use cryogenic coolers for the two higher frequency bands.
Observing requires a two-way link
between the satellite and a  tracking station,
for a wideband down link at 1\,Gbps, and with the uplink used
to transfer a  reference signal.  The current spacecraft mass estimate
is 910\,kg (wet), with a generated power of 1800\,W.

   \begin{table*}
      \caption[]{Comparison of VSOP-2, VSOP and VLBA}
         \label{KapSou}
     $$ 
         \begin{array}{p{0.3\linewidth}rrr}
            \hline
            \noalign{\smallskip}
                  &  VSOP-2 & VSOP & VLBA \\
            \noalign{\smallskip}
            \hline
            \noalign{\smallskip}
Antenna diameter &      9 m &      8 m &  25 m \\
Apogee height    & 25000 km & ~ ~ ~ ~ ~ ~ 21500 km &  0 km \\
Orbital period   &   7.5 hr & 6.3 hr   & 24 hr \\
Polarization     & LCP/RCP  & LCP      & LCP/RCP \\
Data downlink    & 1 Gbps   & 128 Mbps & 512 Mbps \\
Observing frequencies (GHz) & 8, 22, 43 & 1.6, 5 &  ~...1.6, 2, 5, 8, 15, 22, 43, 86 \\
Highest resolution & 38 \mu as & 360 \mu as & 96 \mu as \\
Sensitivity (5/8 GHz) & 22 mJy & 158 mJy & 7.9 mJy \\
Sensitivity (22 GHz) & 39 mJy & ... & 23 mJy \\
Phase-referencing 22 GHz & & & \\
~ ~ (for a 1.5\,hr integration) & 9.1 mJy & ... & 5.3 mJy \\
Launch & 2010 (proposed) & Feb 1997 & ...  \\
            \noalign{\smallskip}
            \hline
         \end{array}
     $$ 
   \end{table*}

\section{New Technical Aspects for VSOP-2}

The on-board radio astronomy antenna is one of the most critical
parts of the spacecraft.
The development of an off-axis mesh antenna with a segmented
(modular) radial rib design has been in progress over the last four
years at ISAS.  
The backup and 
deployment structure is based on the ETS-III project antennas, which are
scheduled to be launched in 2006. To  achieve a 
surface accuracy as high as 0.4\,mm~rms, radial ribs will help shape 
the surface without too
much cable structure. 
HALCA's antenna was designed for use with 22\,GHz as the highest
observing frequency, and there was no adjustment mechanism in
orbit. The VSOP-2 satellite will have in-orbit adjustment mechanism 
with 3 degrees of freedom for main-reflector and 2 degrees of freedom for
the sub-reflector.

The frequency band for the 1\,Gbps VLBI data down-link is 37--38\,GHz, 
and the up-link reference frequency is 40\,GHz. 
The use of both senses 
of polarization will help for the 
maximum use of the allocated band. 
Studies and trade-offs have been
done taking into account both the quantization loss, circuitry complexity,
and down-link power, with the current design employing QPSK
modulation with OFDM (orthogonal frequency division multiplexing)
frequency synthesis.

Phase-referencing observations remove atmospheric phase fluctuations
and consequently can increase the coherence time.
Although this capability was not
considered in the original VSOP mission design, successful
`in-beam' phase-referencing observations have nevertheless
been carried out with
the quasar pairs 1308+326 and 1308+328, separated by 14.3$'$
(Porcas et al.\ \cite{por}) and 1342+662 and 1342+663, separated by 4.8$'$ 
(Guirado et al.\ \cite{gui}).  

Nodding of the whole spacecraft quickly between the calibrator and
target sources is possible with the 
addition of 2 Control Moment Gyroscopes (CMGs) to
the 4 momentum reaction wheels (RWs).
For such phase-referencing observations, orbit determination
accuracy of a few cm is 
required, and this
could be
achieved by adding GPS receivers with a high precision 3-dimensional
accelerometer,
or by using both GPS and Galileo receivers,
according to simulations performed at JPL for the VSOP-2 
orbit.

   \begin{figure*}
   \centering
   \vspace{450pt}
   \includegraphics{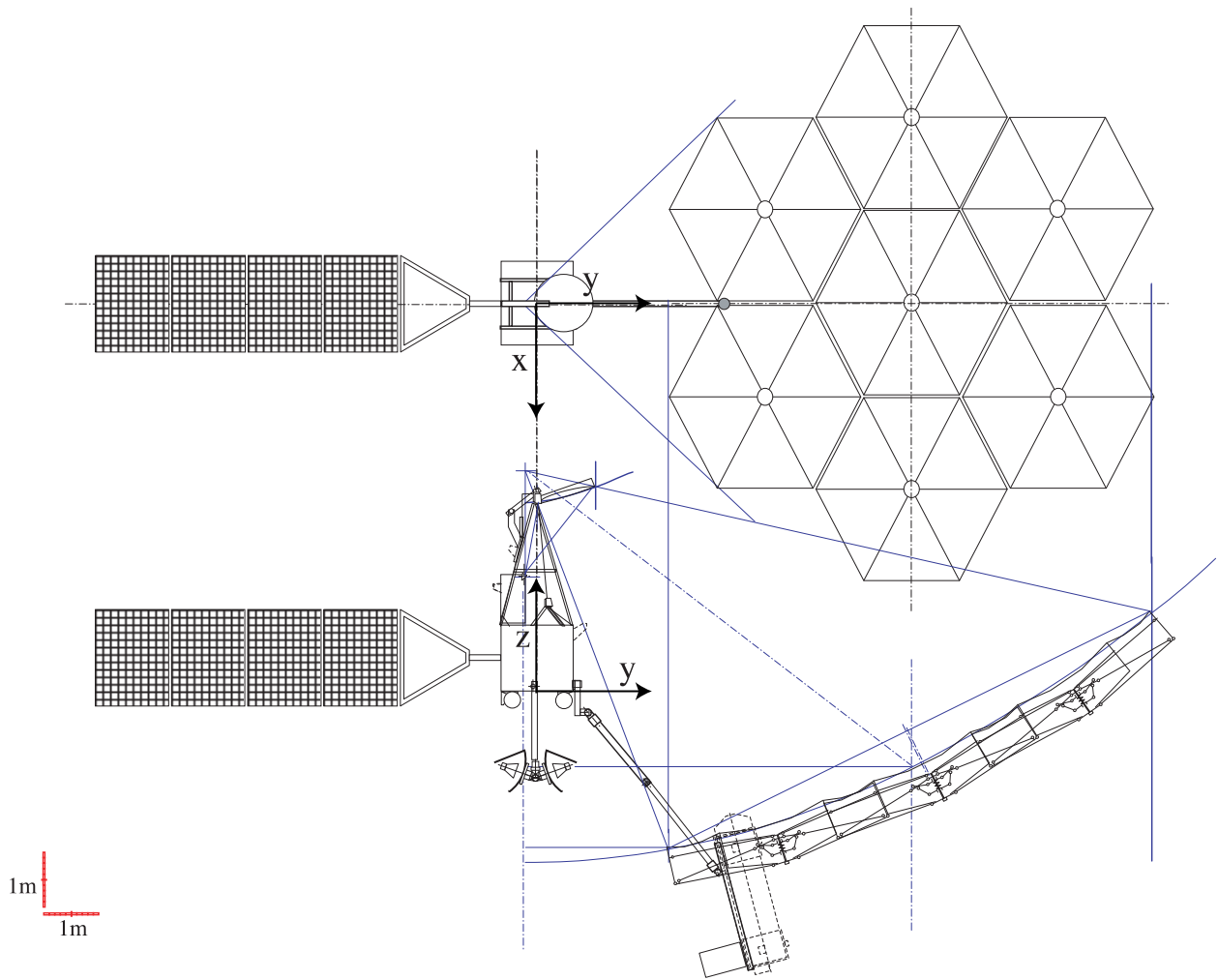}
      \caption{The present design for the 
VSOP-2 spacecraft with a 9\,m diameter offset Cassegrain
antenna of seven hexagonal modules. The
$z$-axis is along the antenna boresight direction.
The main reflector, sub-reflector,
high gain tracking station (``link'') antenna, and solar paddle are 
deployed in orbit. (Two link antennas are shown to illustrate the range
of movement of this single antenna.)
         \label{fig:1}
         }
   \end{figure*}
%

\section{VSOP-2 Proposal Status}

The VSOP-2 proposal was submitted to ISAS in October 2003, at the same
time as the hard X-ray mission NeXT.  Both missions were highly
ranked, but ultimately, due to ISAS' near-term budget profile for the
next few years, neither was included in the budget request for the
2005 fiscal year.  ISAS is preparing a long-term strategic plan
connected with future budget requests.
VSOP-2 will be re-submitted in the next proposal round.

\section{VSOP-2 International Collaboration}

VSOP was realized with large international collaboration in terms of ground 
telescope arrays and correlators, and with VLBI tracking stations.  The
same kind of model is assumed for VSOP-2.  We had rather small-scale 
collaboration from Asian countries for VSOP, but the Japanese
4-station VERA (VLBI Exploration of Radio Astrometry) and the
3-station Korean VLBI Network (KVN) will be capable at 8, 22 and 43\,GHz 
before the VSOP-2 launch.  
The possibility of the Shanghai and Urumqi antennas having
higher frequency upgrades is also being considered, and there is a preliminary
discussion of part-time east-Asian network in the future (Shen et al.\ 2004).
       
Realization of tracking stations network with 
supports from space agencies is important, and one promising possibility is
that a decommissioned  
European antenna can be modified to work as a tracking station.
The ongoing support and assistance from 
international community will be required for the submission 
and realization of the project.

After VSOP-2, we can think of far future missions like mm, sub-mm space 
VLBI, multi-element space mission, etc. VSOP-2 will serve as a precursor 
for these both from scientific and instrumental point of view. 
X-ray interferometry mission in space is under consideration (MAXIM),
and X-ray fringes have been detected in the laboratory. But radio
interferomery is far more advanced, and lots more should be done 
before an X-ray interferometry mission becomes possible.
We hope that 
we can realize this mission to go further to prove exotic phenomena like 
strong general relativistic effects around super-massive black holes, 
and so on.


\end{document}